\documentclass[
 aip,
 amsmath,amssymb,
 reprint,%
author-year,%
]{revtex4-1}

\usepackage{graphicx}
\usepackage{dcolumn}
\usepackage{bm}
\usepackage{float}
\usepackage[utf8]{inputenc}
\usepackage[T1]{fontenc}
\usepackage{mathptmx}
\usepackage{etoolbox}
\usepackage{bbold}
\usepackage{bbm}
\usepackage{txfonts}
\usepackage{xcolor}
\usepackage[colorlinks=true,linkcolor=red,citecolor=blue,urlcolor=cyan]{hyperref}
\graphicspath{ {images/} } 
\makeatletter
\def\@email#1#2{%
 \endgroup
 \patchcmd{\titleblock@produce}
  {\frontmatter@RRAPformat}
  {\frontmatter@RRAPformat{\produce@RRAP{*#1\href{mailto:#2}{#2}}}\frontmatter@RRAPformat}
  {}{}
}%
\makeatother

\renewcommand{\arraystretch}{1.15}

\begin{document}

\preprint{AIP/123-QED}

\title[RKD suprathermal electrons resolved with ALPS]{Temperature anisotropy instabilities of solar wind electrons with regularized Kappa-halos resolved with ALPS}
\author{D. L. Schröder}
\author{H. Fichtner}%
 \email{dustin.schroeder@rub.de}
\affiliation{ 
Institut für Theoretische Physik, Lehrstuhl IV: Plasma-Astroteilchenphysik, Ruhr-Universität Bochum, D-44780 Bochum, Germany
}%

\author{M. Lazar}
\affiliation{%
Centre for mathematical Plasma-Astrophysics, KU Leuven, 3001 Leuven, Belgium
}%

\author{D. Verscharen}
\affiliation{
Mullard Space Science Laboratory, University College London, Dorking RH5 6NT, UK}

\author{K. G. Klein}
\affiliation{%
Department of Planetary Science, University of Arizona, Tucson, USA}%


\date{\today}

\begin{abstract}

Space plasmas in various astrophysical setups can often be both very hot and dilute, making them highly susceptible to waves and fluctuations, which are generally self-generated and maintained by kinetic instabilities. In this sense, we have in-situ observational evidence from the solar wind and planetary environments, which reveal not only wave fluctuations at kinetic scales of electrons and protons, but also non-equilibrium distributions of particle velocities. This paper reports on the progress made in achieving a consistent modeling of the instabilities generated by temperature anisotropy, taking concrete example of those induced by anisotropic electrons, such as, electromagnetic electron-cyclotron (whistler) and firehose instabilities. The effects of the two main electron populations, the quasi-thermal core and the suprathermal halo indicated by the observations, are thus captured. The low-energy core is bi-Maxwellian, and the halo is described for the first time by a regularized (bi-)$\kappa$-distribution (RKD), which was recently introduced to fix inconsistencies of standard $\kappa$-distributions (SKD). In the absence of a analytical RKD dispersion kinetic formalism (involving tedious and laborious derivations), both the dispersion and (in)stability properties are directly solved numerically using the numerical Arbitrary Linear Plasma Solver (ALPS). The results have an increased degree of confidence, considering the successful testing of the ALPS on previous results with established distributions.
\end{abstract}

\maketitle
\textbf{Credits and permissions:} This article may be downloaded for personal use only. Any other use requires prior permission of the author and AIP Publishing. This article appeared in \textit{Phys. Plasmas} 32, 032109 (2025) and may be found at \url{https://doi.org/10.1063/5.0254526}.
\section{\label{sec:level1} Introductory motivation}

Understanding the dynamics of space plasmas, such as the solar wind and planetary environments, presumes modeling particle velocity distributions \citep{Kasper-etal-2002, Stverak-etal-2008, Gary-2015, Wilson-etal-2019a, Wilson-etal-2019b, Lazar-Fichtner-2021b}.
Since these plasmas are hot and sufficiently diluted, the velocity distributions of the particles are often not in thermal equilibrium, as also proven by in-situ observations that show the presence of suprathermal populations and kinetic anisotropies \citep{Verscharen_2019LRSP}.
For the same reasons, collisions are rare, and we expect that the transport of momentum and energy is governed primarily by waves/fluctuations and turbulence \citep{Marsch-2006, Pierrard-etal-2011, Pierrard-Pieters-2014, Yoon-2015}.

Observations consistently report electromagnetic fluctuations at kinetic proton and electron scales within the solar wind and planetary magnetospheres, although their origins are not fully understood \citep{Jian_2009ApJ,Verscharen_2019LRSP}.
Large-scale perturbations from the solar atmosphere's coronal outflows are conveyed by the super-Alfvénic solar wind and decay to smaller scales where dissipation occurs.
Instead, locally generated fluctuations measured in-situ at small proton and electron scales can emerge from kinetic instabilities driven by non-thermal features in particle velocity distribution functions (VDFs), such as temperature anisotropy and particle beams \citep{Stverak-etal-2008, Bale-etal-2009, Wilson2013JGRA, Gary-2015, Gary-etal-2016, Woodham_2019}.
It therefore follows that the implications of these waves and fluctuations can be understood by decoding the wave dispersion and stability properties of the observed non-equilibrium distributions.

To describe these distributions, the more advanced are the $\kappa$-power-law models, which can reproduce the main kinetic anisotropies, but especially the suprathermal populations with enhanced high-energy tails \citep{Maksimovic-etal-2005, Stverak-etal-2008, Lazar_2017_AA, Wilson-etal-2019a, Wilson-etal-2019b, Scherer-etal-2021}.
The $\kappa$-distribution was introduced more than five decades ago as a global empirical model, incorporating not only the suprathermal populations but also the quasithermal core population at low energies \citep{Olbert1968, Vasyliunas1968}.
Later, the modeling of the electron distributions observed in-situ was refined, differentiating between the (bi-)Maxwellian core and the suprathermal halo population reproduced by the (bi-)$\kappa$-distributions \citep{Maksimovic-etal-2005, Stverak-etal-2008, Wilson-etal-2019a}. 
What was revealed was that the core and the halo can have distinct and even opposite anisotropies, e.g., with respect to the interplanetary magnetic field direction \citep{Pierrard-etal-2016}.
These properties are particularly important for the analysis of waves and instabilities, where $\kappa$-distributions have been widely exploited \citep[see reviews by ][]{Helberg-etal-2005, Pierrard2010,Shaaban-etal-2021}. 

More recently, there has been a series of advances in employing these models in a consistent manner, as reported in \cite{Lazar-Fichtner-2021b}.
Remarkable is the adjustment to the so-called regularized $\kappa$-distributions (RKD), \citep{Scherer2018}, for which the values taken by $\kappa$ power exponents are no longer restricted; see also the detailed discussion in Section~II below. 
The moments of the standard $\kappa$-distribution (SKD) and the corresponding transport coefficients are not well-defined for all values of $\kappa$ \citep{Scherer2018}. The RKD is a novel approach to alleviate this shortcoming of the SKD. For RKDs,  the moments are well-defined for all $\kappa>0$. Compared to the SKD, however, the RKD carries increased mathematical complexity which precludes the analytical evaluation of the general linear dispersion relation in systems with an RKD background \citep{Scherer_2019, Lazar-etal-2020, Husidic-etal-2022}.
Attempts to derive dielectric response of plasma electrons with RKD are so far only known for longitudinal electrostatic waves \citep{Scherer2018, Gaelzer-etal-2024}. 
The derivation of the dielectric properties is even more difficult in the case of magnetized plasmas with anisotropic RKD distributions. 

Therefore, in this paper we motivate the use of the Arbitrary Linear Plasma Solver (ALPS) \citep{Verscharen_2018}, for a direct numerical evaluation of the dispersion and stability properties of RKD plasmas. 
We address a series of instabilities driven by temperature anisotropy of electron populations, which are often invoked to explain their properties, in particular the anisotropy limitations revealed by in situ observations \citep{Stverak-etal-2008, Xu-Chen-2012, Lazar-etal-2017, Shaaban-etal-2019, Yoon-etal-2024}.
Thus, for a temperature excess in the direction perpendicular to the magnetic field, $A = T_\perp/T_\parallel >1$ (where $\parallel, \perp$ are directions with respect to local magnetic field), linear theory predicts instabilities of electromagnetic (EM) cyclotron modes, while for an opposite anisotropy, $A <1$, the kinetic firehose instabilities can be excited.
The present work restricts to the instabilities with wavevectors that are parallel to the background magnetic field, which are oscillatory (with finite wave frequency) and often prove sufficiently effective in competition with the oblique (aperiodic) excitations \citep{Gary_2006JGRA, Shaaban-etal-2019, Sarfraz-etal-2022}.
\textcolor{black}{\cite{Husidic_2020} analyzed the same instabilities using a global bi-RKD representation incorporating both the core and halo populations, which they resolved with LEOPARD \citep{Astfalk-Jenko-2017}, another solver for arbitrary velocity distributions.
Instead, here we adopt a more complex but also more realistic dual distribution of the electron populations, describing accordingly to the observations a bi-Maxwellian core and, for the first time, a bi-RKD halo.}
The test cases and the new results obtained with ALPS are very promising, offering perspectives for extended analysis of distribution models of even greater complexity.

The structure of the paper is as follows. We begin with a brief summary of the theory of dual core-halo modeling in Section \ref{sec:Theory_CH_Model} and of the numerics of wave instability in the solver ALPS in Section \ref{sec:Numerics_ALPS}. This is followed with validating the ALPS implementation against a series of previous results, particularly those obtained by \citet{Lazar_2017_AA} for parallel modes in plasmas with Maxwellian core and $\kappa$-halo in Section \ref{sec:Validation}, with focus on EMEC in Section \ref{sec:Validating EMEC} and on EFHI in part \ref{sec:Validating EFHI}. The new results are presented and discussed in the first and second parts of Section \ref{sec:RKD results}. All findings are summarized in the concluding Section \ref{sec:conclusion}.

\section{Consistent dual distributions: Maxwellian core plus Regularized $\kappa$- halo}\label{sec:Theory_CH_Model}

$\kappa$-modelling is extensively utilized for diagnosing space plasmas and conducting analysis of their kinetic properties, such as \cite{Pierrard2010} or \cite{lazar2022kappa}. The $\kappa$-distribution function resembles a Maxwellian distribution at low energies but transitions to a power-law at higher energies. 
This power-law feature allows for fitting the observed high-energy tails of the solar wind electron distribution, enhanced by the so-called halo component, while representing the core with a Maxwellian \citep{Maksimovic-etal-2005, Stverak-etal-2008}.
In fact, the $\kappa$-model was introduced as a global distribution incorporating both the core and halo populations \citep{Olbert1968, Vasyliunas1968, Maksimovic-etal-1997}. This way it simplifies the analysis by reducing the number of parameters, and makes it easier to handle in both observational and theoretical studies. 
However, a global $\kappa$ does not always provide an accurate fit to the observed distributions, forcing the core and suprathermal populations to share the same parameters, such as density, temperature, and anisotropy, which is not fully justified \citep{Maksimovic-etal-2005, Stverak-etal-2008, lazar2022kappa}.

More sophisticated or composed models can reproduce multiple components with different properties, and generally provide more accurate details of the observed distributions \citep{Maksimovic-etal-2005, Wilson-etal-2019a, Wilson-etal-2019b}. For instance, at large enough heliospheric distances (e.g., $>0.6$~AU), or in general in slow solar wind, the beaming (or strahl) electrons are much less dense than the core and halo (\(n_s << n_h < n_c\), with \( n_{c} \) and \( n_{h} \) the number densities of the core and halo components and $n_s$ the number density of the strahl component), and the observed distributions can be, when neglecting the strahl component, better described by a dual Maxwellian-$\kappa$-model. 
This usually includes a bi-Maxwellian for the core (subscript $c$) at low energies, and a bi-$\kappa$ for the suprathermal halo (subscript $h$) \citep{Maksimovic-etal-2005, Stverak-etal-2008, Lazar_2017_AA}:
\begin{equation}\label{eq:core-halo model}
f_{}(v_{\parallel}, v_{\perp}) = \dfrac{n_c}{n} f_{c}(v_{\parallel}, v_{\perp}) +\dfrac{n_h}{n} f_{h}(v_{\parallel}, v_{\perp}).
\end{equation}
$v_{\parallel,\perp}$ is the particle velocity parallel and perpendicular to the magnetic background field and \( n = n_c + n_h\) represents the total number density of plasma particles.
Such a dual model can then reproduce different (even opposite) anisotropies of the core and halo, as indicated by various events \citep{Pierrard-etal-2016}.

Thermal particle populations in a magnetized plasma, in particular those  with anisotropic temperatures (with respect to the uniform magnetic field direction), are commonly be described by a bi-Maxwellian distribution 
\begin{equation}\label{eq: fmaxwell}
f_M(v_{\|}, v_{\perp})= N_M \exp{\left(- \frac{v_{\|}^2}{\theta_{M, \|}^2}- \frac{v_{\perp}^2}{\theta_{M,\perp}^2}\right)}.
\end{equation}
$\theta_{M \|, \perp} = \sqrt{2k_B T^M_{\|, \perp}/m}$, with the Boltzmann constant $k_B$, denotes the thermal speed of particles with mass $m$, associated with the Maxwellian temperature $T^M_{\|, \perp}$ ($\parallel, \perp$ denoting directions with respect to the magnetic field). The normalization factor is given by $N_M=1/(\pi^{3/2} \theta_{M,\|} \theta_{M, \perp}^2 )$.
Early studies have ignored the effects of suprathermal halo populations, and widely described kinetic instabilities driven by anisotropic temperatures, i.e., $A = T_\perp / T_\parallel \ne 1$, limited to these idealized bi-Maxwellian distributions \citep{Gary-1993}.

Later studies have generalized the investigations of kinetic instabilities by adopting a bi-$\kappa$-distribution, hereafter called \textit{standard $\kappa$-distribution}, also SKD for short \citep{Lazar_Fichtner_Yoon_2016}
\begin{equation}\label{eq: f_skd}
f_{\mathrm{SKD}}\left(v_{\|}, v_{\perp}\right)=N_{\mathrm{SKD}}\left(1+\frac{v_{\|}^2}{\kappa \Theta_{\|}^2}+\frac{v_{\perp}^2}{\kappa \Theta_{\perp}^2}\right)^{-\kappa-1}
\end{equation}
with normalization constant 
\begin{equation}
N_{\mathrm{SKD}}=\frac{1}{\pi^{3 / 2} \Theta_{\|} \Theta_{\perp}^2} \frac{\Gamma(\kappa+1)}{\kappa^{3 / 2} \Gamma(\kappa-1 / 2)}.
\end{equation}
The parameter $\kappa$ is a dimensionless positive real number, and $\Gamma$ represents the Gamma function.
The thermal speeds are $\Theta_{\|,\perp}=\sqrt{2k_BT^{\kappa}_{\|,\perp}/m}$ with $T^{\kappa}_{\|,\perp}=\dfrac{\kappa}{\kappa-3/2}T^{M}_{\|,\perp}>T^{M}_{\|,\perp}$.
If the SKD is adopted as a global model, the highlighting of the new effects of the suprathermal populations can be done by contrast with those previously obtained counting only on the bi-Maxwellian core. 
However, the latter can be obtained directly in the limit $\kappa \to \infty$, when the $\Theta_{\parallel, \perp}$ parameters approximate the core thermal velocities $\Theta_{\parallel, \perp} \to \theta_{M\parallel, M\perp}$ and are therefore independent of $\kappa$ \citep{Lazar-etal-2015, Lazar_Fichtner_Yoon_2016}.

\begin{figure*}[t!]
    \centering
    \includegraphics[width=0.45\textwidth]{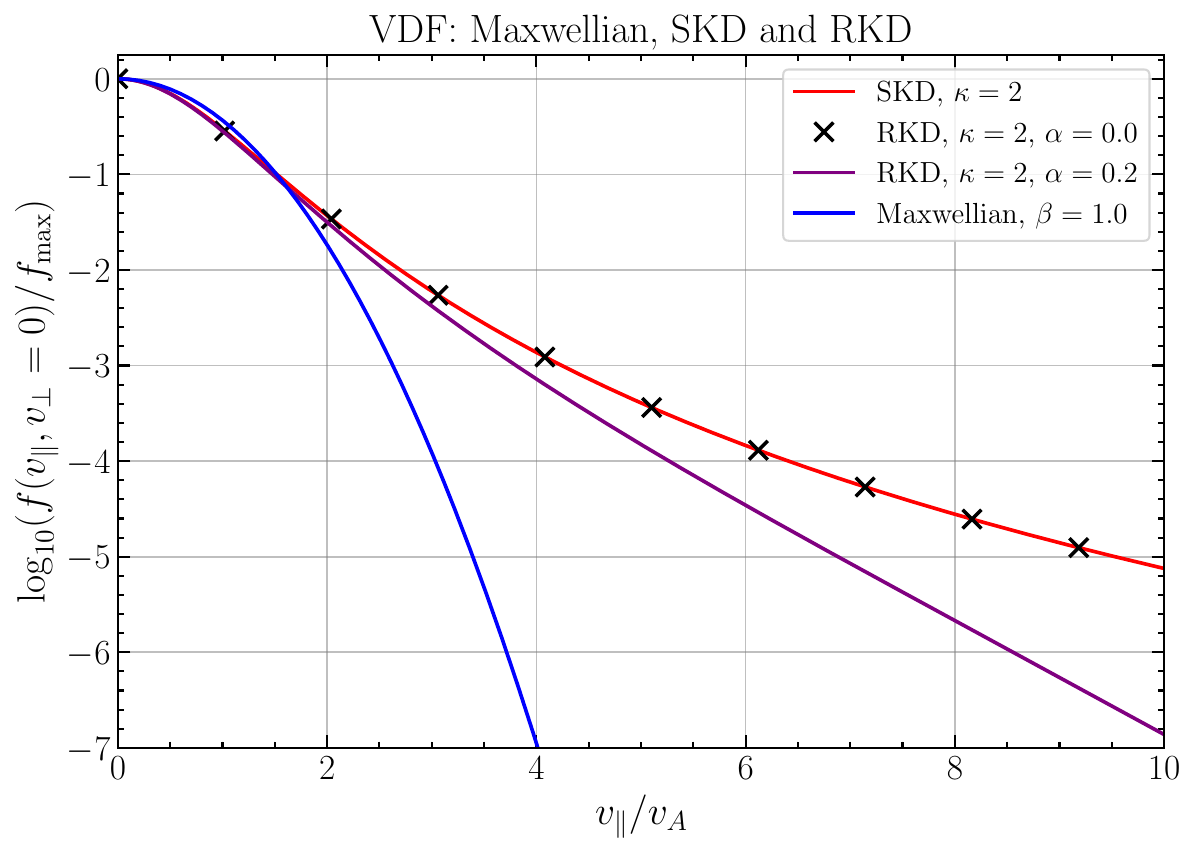}
    \hspace{0.05cm} \includegraphics[width=0.45\textwidth]{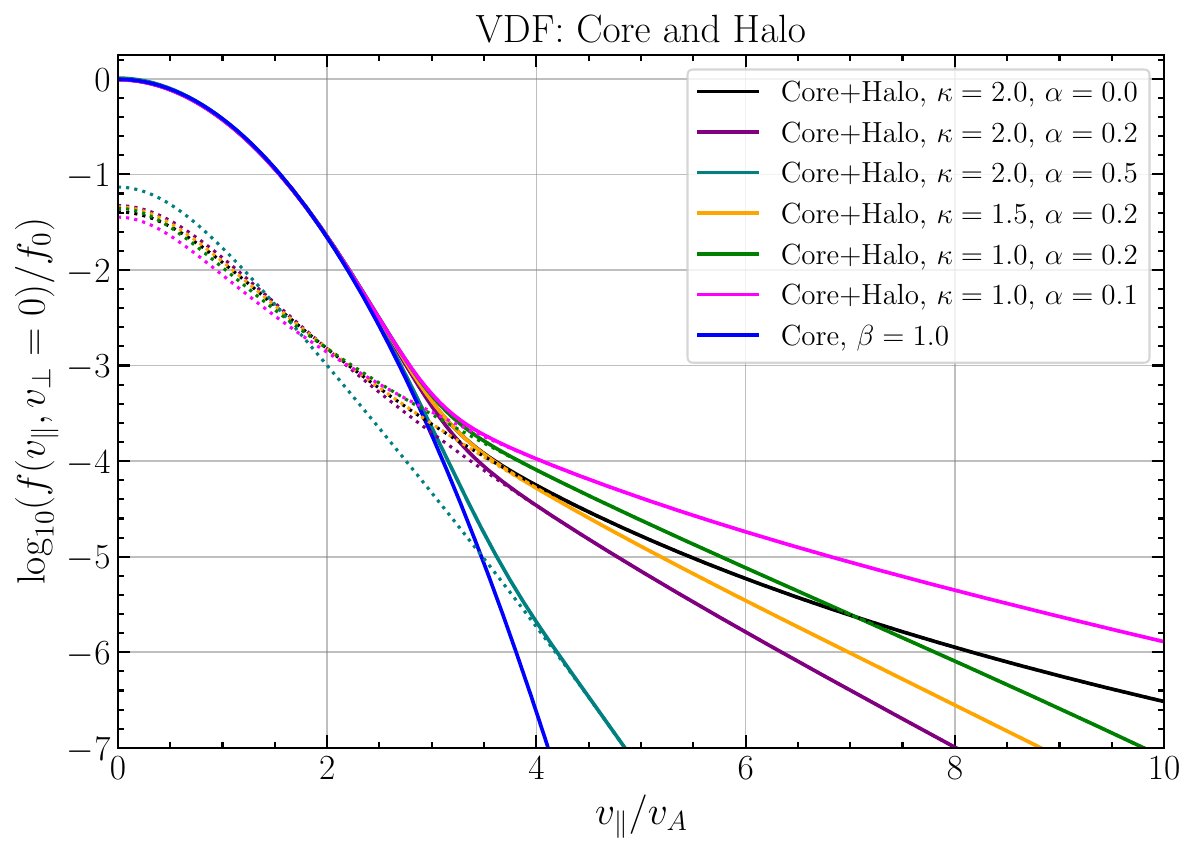}
    \caption{\textit{In the left panel a comparison between SKD (red) and RKD, with different $\kappa$ and $\alpha$ values, and Maxwellian (blue), normalized to their maxima, with the same plasma beta of $\beta =1.0$. When $\alpha \rightarrow 0$ (black cross), the RKD approaches the SKD, while for an increasing $\alpha$, the Maxwellian case is reached. With a decreasing $\kappa$, the high energy tails become more prominent. In the right panel a comparison between the different anisotropic core+halo VDFs (solid) and the Maxwellian Core (blue) with a density contrast $n_h/n_c=0.05$.  The RKD/SKD-limit halos are presented with a dotted line. This represents our used core-halo model. }}
    \label{Plot_Maxwell_SKD_comparison_shifted}
\end{figure*}

From kinetic theory, the magnetohydrodynamic equations of fluid theory can be derived by forming \textcolor{black}{velocity moments
$M_l$ of the $l$-th order, where $l \in{\{0, 1, 2,...\}}$ \cite[see general definitions for scalar, vector and tensor moments in][]{Scherer2018}.} All moments should exist, meaning that all $M_l$ converge. For a Maxwellian distribution, this condition is satisfied. However, for an SKD, $M_l < \infty$ only for $l<2\kappa -1$ \citep{Scherer2018}. To provide a meaningful description of the plasma, a kinetic temperature must exist, defined in fluid theory through the second velocity moment. If one requires the second moment $l=2$ to be well-defined, the condition $l<2\kappa-1$ demands that $\kappa>3/2$. However, observations have identified space plasmas with $\kappa \leq 3/2$ \citep[see, e.g.,][]{Gloeckler-etal-2012}.
To address these unphysical properties and the issue of divergent moments that hinder a closed description of a physical system and imply undesired limitations of the $\kappa$-parameter, the regularized $\kappa$-distribution (RKD) $f_{RKD}$ was introduced \citep{Scherer2018}:
\begin{equation}\label{eq: frkd}
\begin{aligned}
f_{\mathrm{RKD}}\left(v_{\|}, v_{\perp}, \alpha \right)= &  N_{\mathrm{RKD}}\left(1+\frac{v_{\|}^2}{\kappa \Theta_{\|}^2}+\frac{v_{\perp}^2}{\kappa \Theta_{\perp}^2}\right)^{-\kappa-1} \\
& \times \exp \left(-\frac{\alpha_{}^2 v_{\|}^2}{\Theta_{\|}^2}-\frac{\alpha_{}^2 v_{\perp}^2}{\Theta_{\perp}^2}\right)
\end{aligned}
\end{equation}

$N_{RKD}$ is a normalization factor, given by 
\begin{equation}
N_{\mathrm{RKD}}=\frac{1}{\pi^{3 / 2} \Theta_{\|} \Theta_{\perp}^2 W}  .
\end{equation}
where 
\begin{equation}
    W = U\left(\frac{3}{2}, \frac{3-2\kappa}{2},\alpha^2 \kappa \right)
\end{equation}
and $U$ denotes the Tricomi function \citep{Scherer_2019}.
Also it is possible to consider more general distributions with a direction-dependent cut-off parameter \citep{Scherer-etal-2020}. The dimensionless cut-off parameter $\alpha$ controls the strength of the exponential decay and is independent of $\kappa$. The idea of the RKD is to combine the SKD with an exponential Maxwellian-like part, to dampen the tail of the distribution, preventing the divergences associated with the SKD ensuring the existence of all velocity moments. For a relativistic generalization of the RKD, see \citet{HanThanh-etal-2022}.


As shown in Fig.~\ref{Plot_Maxwell_SKD_comparison_shifted}, left panel, the RKD's power-law component dominates at intermediate velocities compared to the exponential function, reflecting the behavior of suprathermal particles, while the latter dominates at very high velocities, resulting in the desired cut-off. The significant advantage of the RKD is that all velocity moments,
\begin{equation}
\begin{aligned}
M_{l}(\kappa, \alpha, \Theta) 
&\equiv  N_{RKD} I(\kappa, \alpha, l, \Theta), 
\end{aligned}
\end{equation}
\textcolor{black}{are well defined for all values of $\kappa > 0$ compared to SKD \citep{Lazar_Fichtner_Yoon_2016} or other similar attempts, e.g., \cite{Shahzad-etal-2024}.
The integral $I(\kappa, \alpha, l, \Theta)$ becomes analytically solvable for all $l \in \{0, 1, 2, ...\}$ and does not diverge, making all velocity moments calculable; 
see the analytical expressions derived in \cite{Scherer_2019}, including for anisotropic RKDs with temperature anisotropies and skewed or drifting distributions of beam-plasma systems.} 
Therefore, an unphysical limitation of $\kappa$ is no longer necessary, and the temperature is always well-defined as the second moment of the RKD. The RKD is thus defined for any $\kappa > 0$, enabling a closed description of a physical system using fluid equations at a macroscopic level. Furthermore, the RKD includes an SKD and a Maxwellian distribution as limiting cases, as shown in Fig.~\ref{Plot_Maxwell_SKD_comparison_shifted}, left panel. In the limit $\alpha\rightarrow 0$, the RKD recovers the SKD from Eq. (\ref{eq: f_skd}). In the limit $\kappa\rightarrow \infty$ and $\alpha\rightarrow 0$, the RKD recovers the Maxwellian distribution from Eq. (\ref{eq: fmaxwell}) under a suitable choice of $\Theta$.
In Fig.~\ref{Plot_Maxwell_SKD_comparison_shifted}, right panel, the dual-core-halo model is plotted: The Maxwellian core, with a higher density, combined with a RKD halo with lower density.

The RKD has been successfully applied in practice, for example, in macroscopic parameterizations of suprathermal populations \citep{Lazar-etal-2020}, modeling of anisotropic distributions measured in situ in the solar wind \citep{Scherer-etal-2021}, evaluation of transport coefficients in non-equilibrium plasmas \citep{Husidic-etal-2022}, or Harris sheet equilibrium modeling \citep{Hau-etal-2023}. Additionally, when $\alpha$ is chosen small enough, the cut-off only occurs for velocities outside the measurement range, and an RKD fit applied to measurement data provides the same results as those previously obtained through an SKD, eliminating the need for extensive conversion steps \citep{lazar2022kappa}. 
The RKD retains the flexibility of the $\kappa$- distribution while ensuring that all velocity moments converge. The RKD is thus well-suited for describing space plasmas where nonthermal features are significant. Overall, the RKD provides a physically consistent representation and interpretation of suprathermal particles.

The general dispersion tensor for RKDs has not been derived yet, but the investigation of these modes is possible with ALPS \citep{Verscharen_2018,ALPS:2023}. Made public in 2023, this numerical solver directly evaluates the linear Vlasov-Maxwell dispersion relation in a plasma with arbitrary gyrotropic background VDFs. Additionally, ALPS addresses irregularities and challenges encountered with previous similar solvers \citep{Astfalk-Jenko-2017, Husidic_2020}.

\section{Wave instability in numerics: ALPS}\label{sec:Numerics_ALPS}

We assume that the plasma fluctuations, specifically those of the electric and magnetic fields (\(\boldsymbol{E}\) and \(\boldsymbol{B}\)) and those in the VDF, are small enough to justify the application of linear theory. To investigate plasma instabilities, one solves the kinetic dispersion relation to obtain the complex frequency
\begin{equation}
    \omega(\boldsymbol{k}) = \omega_r + i\gamma,
\end{equation}
where \(\boldsymbol{k}\) is the wave vector, and \(\omega_r\) and \(\gamma\) denote the real and imaginary parts of the frequency, respectively.

To obtain this solution, one uses the linearized Vlasov equation, providing an expression for the plasma susceptibilities \(\boldsymbol{\chi}_j\) for the \(j\)-th species (see Appendix \ref{app_disper}). These susceptibilities are then related to the plasma's dielectric tensor \(\boldsymbol{\epsilon}\) through
\begin{equation}
    \boldsymbol{\epsilon} = \mathbbm{1} + \sum_j \boldsymbol{\chi}_j.
\end{equation}
with the unity tensor $\mathbbm{1}$.

From this, one can derive
\begin{equation} \label{alps_eq}
    \boldsymbol{n} \times (\boldsymbol{n} \times \boldsymbol{E}) + \boldsymbol{\varepsilon} \cdot \boldsymbol{E} \equiv \mathcal{D} \cdot \boldsymbol{E} = 0,
\end{equation}
where \(\boldsymbol{n} = \boldsymbol{k}c/\omega\) and \(c\) is the speed of light. Solving \(\text{det}\ \mathcal{D} = 0\) provides non-trivial solutions to Eq. (\ref{alps_eq}) in terms of $\omega(\boldsymbol{k})$ which are considered the solutions to the dispersion relation. The ALPS code \citep{Verscharen_2018} will be used to obtain these solutions.

ALPS, a parallelized numerical code written in Fortran-90, is designed to solve Eq. (\ref{alps_eq}) for various plasma conditions, including hot non-relativistic and relativistic plasmas. The code's versatility allows for the examination of an arbitrary number of species with equilibrium distribution functions $f_{0j}$, accommodating arbitrary propagation directions with respect to the undisturbed magnetic field (referred to as the 'background field').

To utilize ALPS, users need to input numerical values for $f_{0,j}(p_{\perp},p_{\|})$, where $p_{\|,\perp}$ denote the parallel and perpendicular components of momentum relative to the background field. These values can be organized into an ASCII table. Additionally, initial guesses for $\omega_r$ and $\gamma$ must be provided.

The code employs an iterative Newton-secant method to solve Eq. (\ref{alps_eq}), resulting in the determination of $\omega_r(\boldsymbol{k})$ and $\gamma(\boldsymbol{k})$. For more details on its implementation and capabilities, please refer to \citet{Verscharen_2018}.



\section{Validation of solver ALPS}\label{sec:Validation}

To begin with, the ALPS setup is validated against results from \cite{Lazar_2017_AA}, who solved Eq. (\ref{eq:disp_emec}) and Eq. (\ref{eq:disp_efhi}) via Mathematica. This validation extends earlier ones presented, e.g., in \citet{Verscharen_2018}. 
The VDFs are represented as a dual Maxwellian-$\kappa$-model in ALPS by using Eq (\ref{eq:core-halo model}) with $f_c = f_M$ from Eq  (\ref{eq: fmaxwell}) and $f_h = f_{SKD}$ from Eq (\ref{eq: f_skd}). The corresponding input VDFs are shown in Fig. \ref{Plot_Maxwell_SKD_comparison_shifted}, right panel.

The plasma frequency is $\omega_{h, j} = \sqrt{4 \pi n_h q_j^2 / m_j}$, while $\Omega_{ j} = q_j B_0 / (m_j c)$ is the non-relativistic gyrofrequency and $v_{A, \text{ref}}/c= v_{A, j}/c= B_0/(\sqrt{4\pi n_{j} m_{j}}c)$ is the reference Alfv\'{e}n speed, where $m_j$ is the species rest mass, $q_j$ is the charge of the species, $n_j$ is the density of the species and the index $j$ refers to the species, i.e. $j=p$ for protons and $j=e$ for electrons. 
We use a parallel plasma beta of $\beta_{j,c}=8\pi n_ck_bT_{j,c,\|}/B_0^2$ and $\beta_{j,h}=8\pi n_hk_bT_{j,h,\|}/B_0^2$.
Since it has been demonstrated in numerous other studies that an SKD with a kappa-dependent temperature $T_{j, h}=T_{j,h}(\kappa)$ is the more natural choice, we will only use those VDFs for our studies.

\subsection{Electromagnetic electron cyclotron (EMEC) instability}\label{sec:Validating EMEC}

    

\textcolor{black}{Since whistler or EMEC waves propagate at frequencies with $\omega_r \gg \Omega_p$ and have a right-hand circular polarization, their interaction with protons is negligible}, and hence the protons are described with an isotropic Maxwellian.
Since the halo in the solar wind plasma tends to be more anisotropic, hotter, and less dense than the core, the latter is assumed to be isotropic. There may also be cases where both the core and the halo have isotropic populations, but as shown in \cite{Lazar_2018_JGRA}, in such instances, the (Maxwellian) core predominantly drive the instability, resulting in a negligible influence of the halo. This would then limit the study of the influence of different RKD halos \citep{Maksimovic-etal-2005}.   
A scenario with a low plasma beta combined with a high anisotropy and a high plasma beta combined with a low anisotropy is studied.

For the first case, the parameters are chosen as $\beta_{ e, c}=1$, $\beta_{ e, h}=0.05$ and $T_{\perp, e, c}/T_{\|, e, c}=1$ and $T_{\perp, e, h}/T_{\|, e, h}=3$ with $n_c/n_{\text{}} = 0.9523$ and $n_h/n_{\text{}} = 0.0477$, which means a core-halo density contrast of $\eta = n_h/n_c = 0.05$. 
For case 2, the parameters are: $\beta_{ e, c}=1$, $\beta_{ e, h}=1$ and $T_{\perp, e, c}/T_{\|, e, c}=1.1$ and $T_{\perp, e, h}/T_{\|, e, h}=1$ with $n_c/n_{\text{}} = 0.9523$ and $n_h/n_{\text{}} = 0.0477$, which represents a core-halo density contrast of $\eta =n_h/n_c= 0.05$. 
For a detailed interpretation of these EMEC cases, see \cite{Lazar_2017_AA}.

\begin{figure*}[t!]
    \centering
    \hspace{0.3cm}
     \includegraphics[width=0.99\textwidth]{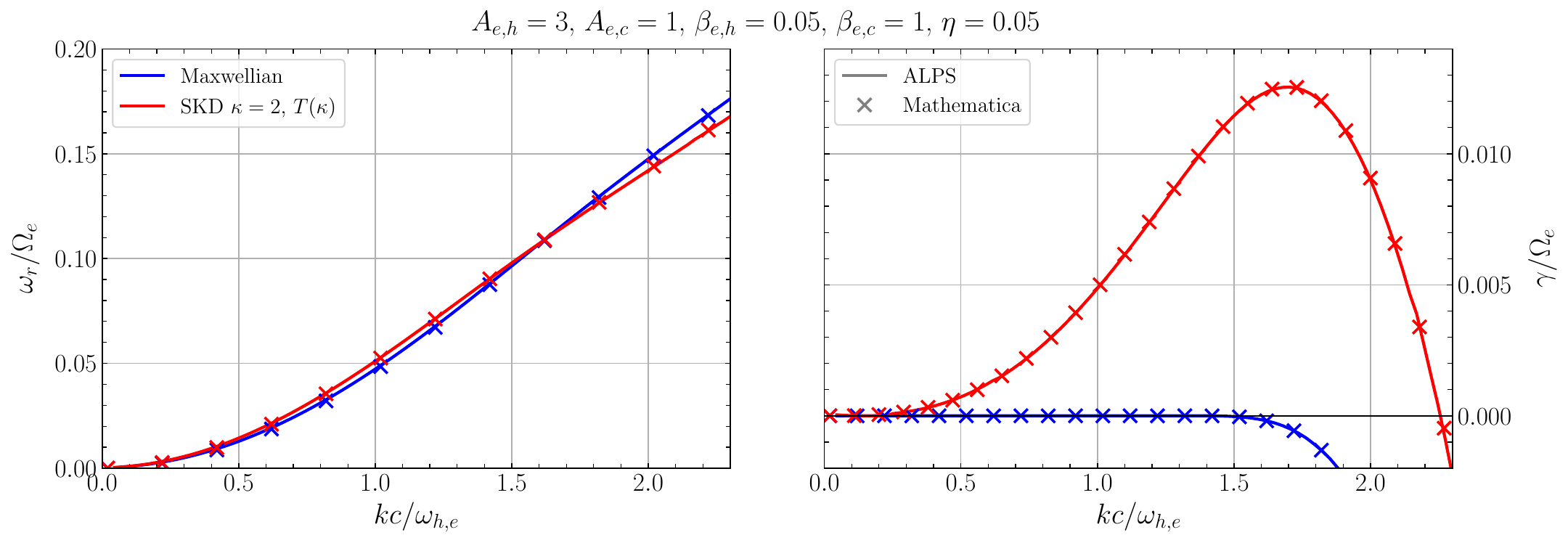}
    \caption{\textit{The EMEC Instability (low halo plasma beta, high halo anisotropy) with different SKD halos (blue: Maxwellian limit, red: SKD, solid: results derived with ALPS, crosses: Mathematica results). The frequency is shown on the left, while growth rate is on the right side. All other parameters are stated above the panels.  }}
    \label{Lazar2017_CH_EMEC_case1}
\end{figure*}

\begin{figure*}[t!]
    \centering
    \hspace{0.3cm}
     \includegraphics[width=0.99\textwidth]{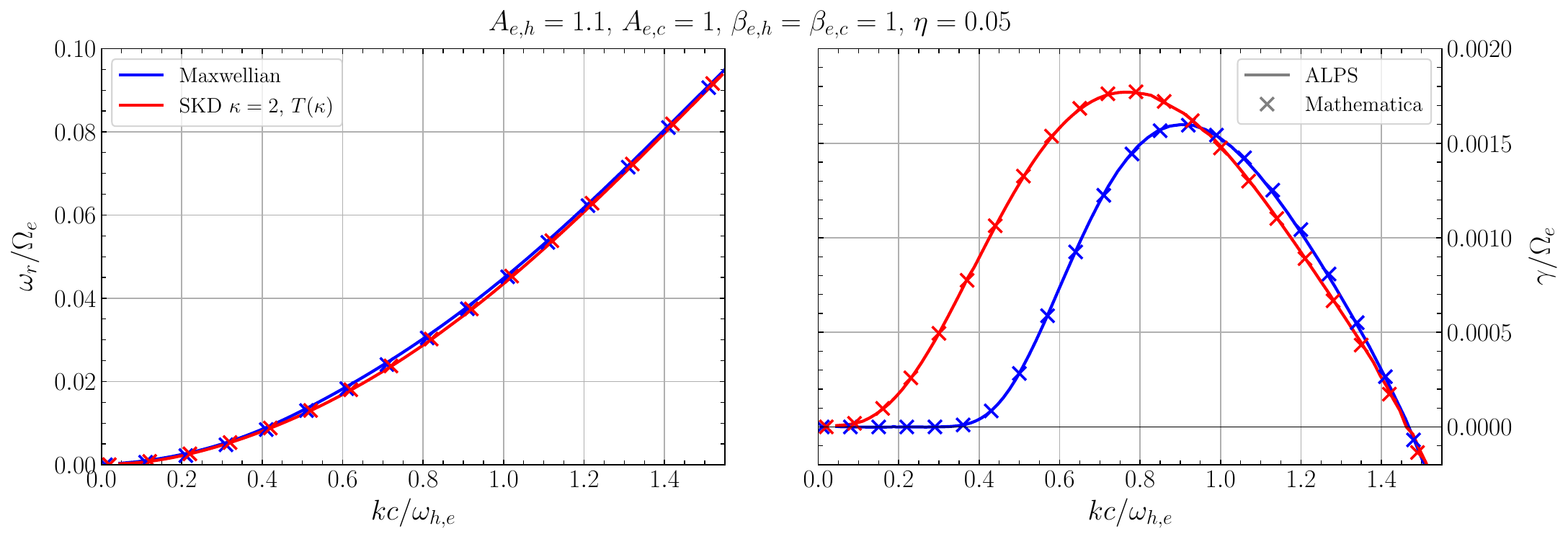}
    \caption{\textit{The EMEC Instability (high halo plasma beta, low halo anisotropy) with different SKD halos (blue: Maxwellian limit, red: SKD, solid: results derived with ALPS, crosses: Mathematica results). The frequency is shown on the left, while growth rate is on the right side. All other parameters are stated above the panels. }}
    \label{Lazar2017_CH_EMEC_case2}
\end{figure*}

The resulting dispersion curves are presented in Fig. \ref{Lazar2017_CH_EMEC_case1} and Fig. \ref{Lazar2017_CH_EMEC_case2}. Shown is the frequency on the left side and growth rate on the right side, with Maxwellian cases in blue and the SKD cases in red.  The solid lines represent the ALPS results, compared to those obtained with Eq. (\ref{eq:disp_emec}) with Mathematica, plotted with crosses.
The agreement for both the frequency and the growth rate for the Maxwellian distribution and the SKD is excellent for both cases, validating not only the solver ALPS itself, but also the implementation of our core-halo model, solidifying the derived RKD results in the following section.     

\subsection{Electron Firehose Instability (EFHI)}\label{sec:Validating EFHI}
Furthermore, the ALPS setup is tested against similar results from \cite{Lazar_2017_AA} for the parallel EFHI. As before, the electrons are assumed to be a dual-core-halo plasma, with the following parameters:  $\beta_{ e, c}=1$, $\beta_{ e, h}=4$ and $T_{\perp, e, c}/T_{\|, e, c}=1$ and $T_{\perp, e, h}/T_{\|, e, h}=0.6$  with $n_c/n_{\text{}} = 0.9523$ and $n_h/n_{\text{}} = 0.0477$, $\eta = 0.05$. The protons need to be described with a dual-core-halo model too, since at typical EFHI frequencies, they can interact with the electrons. For the protons, an isotropic core $T_{\perp, p, c}/T_{\|, p, c}=1.0$ and isotropic halo $T_{\perp, p, h}/T_{\|, p, h}=1.0$ with $\beta_{ p, c}=1$, $\beta_{ p, h}=4$ will be used.

\begin{figure*}[t!]
    \centering
    \hspace{0.3cm}
     \includegraphics[width=0.99\textwidth]{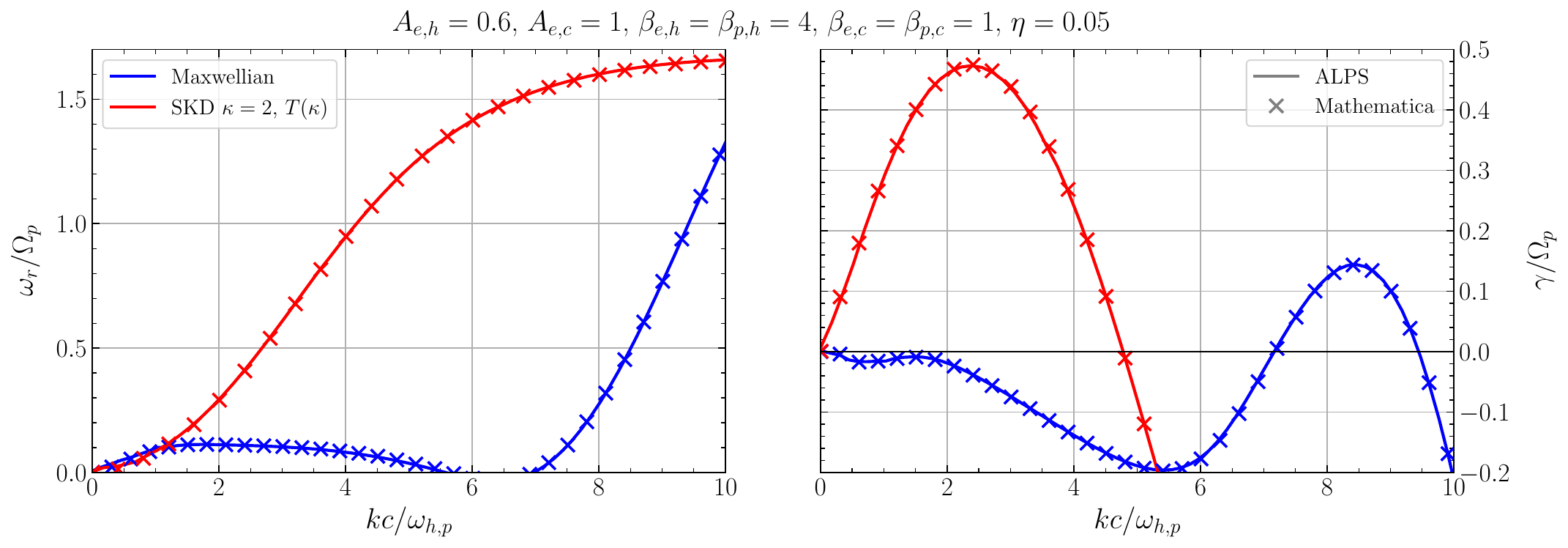}
    \caption{\textit{The EFHI with different SKD halos (blue: Maxwellain limit, red: SKD with kappa dependent temperature, solid: results derived with ALPS, crosses: Mathematica results). The frequency is shown on the left, while growth rate is on the right side. All other parameters are stated above the panels.  }}
    \label{Lazar2017_CH_EFHI_case1}
\end{figure*}

In Fig.~\ref{Lazar2017_CH_EFHI_case1}, the dispersion curves for the EFHI, with frequency on the left-hand side and growth rate on the right-hand side are shown, with the ALPS results plotted with solid lines, compared to the ones derived with Mathematica by solving Eq. (\ref{eq:disp_efhi}) derived in \cite{Lazar_2017_AA}, represented with crosses. 
The computations are again performed for a Maxwellian (blue) and a SKD with $\kappa = 2.0$ (red). The agreement is also excellent. Note that the results for the Maxwellian case differ from those presented in \cite{Lazar_2017_AA}, likely due to differences in earlier versions of Mathematica.
    

\section{Unstable solutions with anisotropic RKD Halo}\label{sec:RKD results}

\begin{figure*}[t]
    \centering
    \includegraphics[width=0.99\textwidth]{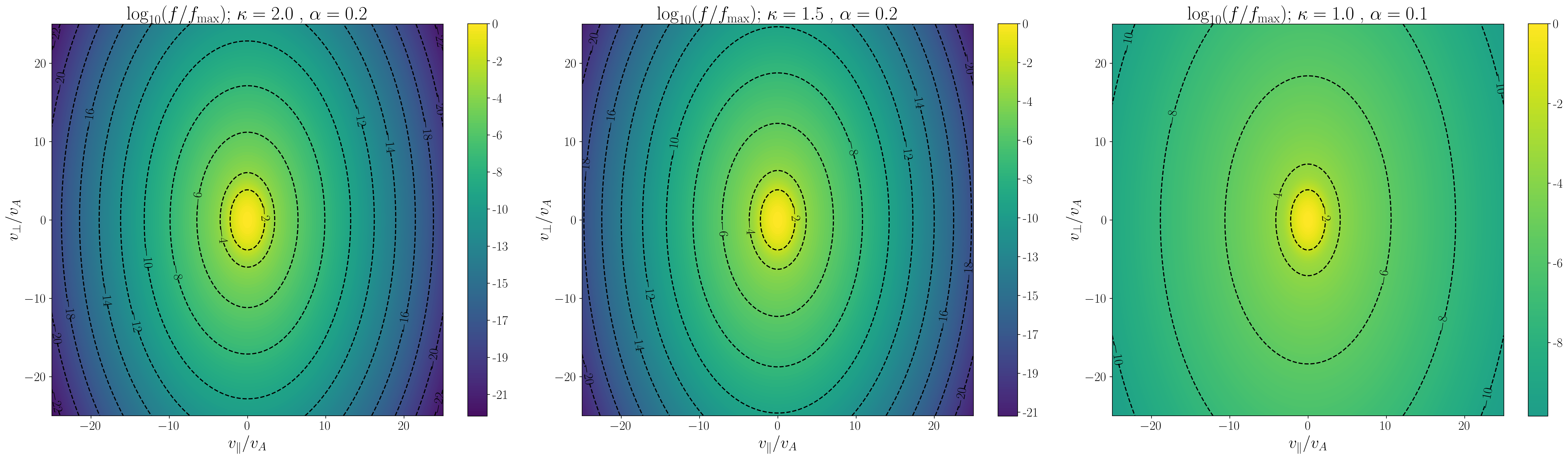}
    \caption{\textit{Contour plots of the different anisotopic RKD Core-Halo VDFs normalized to their maximums. With $\kappa= (2, 1.5, 1)$ from left to right and with $\alpha = 0.2$ for the first and second plot and $\alpha = 0.1$ for the third one . The anisotropy in parallel directions is clearly visible. A lower $\kappa$-value leads to a less steeper decrease of the distribution function. Note that ALPS uses a cylindrical coordinate system, where $v_{\perp} \ge0$, however, for illustrative purposes, negative $v_{\perp}$ values are also shown.}} \label{contourplot_Halo_VDF_normalized}
\end{figure*}

We model the electron halo population with an anisotropic regularized bi-$\kappa$-distribution, as defined in Eq.~\eqref{eq: frkd}. 
Examples of anisotropic core-halo RKDs, used in ALPS, are displayed in Fig. \ref{plot_Halo_VDF_ALPSlike} and Fig. \ref{contourplot_Halo_VDF_normalized}.
\textcolor{black}{While Fig. \ref{plot_Halo_VDF_ALPSlike} is in principal the same as Fig. \ref{Plot_Maxwell_SKD_comparison_shifted}, but a more closer representation of the VDFs that are implemented in ALPS, Fig. \ref{contourplot_Halo_VDF_normalized} shows contour plots of $f=(n_c/n)f_c + (n_h/n)f_h$ for three different values of $\kappa$ } (from left to right $(2, 1.5, 1.0)$) with the same $\alpha = 0.2$, normalized to their maxima. One can see the anisotropy in parallel direction and the effect of lower $\kappa$-values, i.e.: the decrease of $f$ for $\kappa = 1.0$ with increasing velocity is noticable less steeper than for $\kappa = 2.0$. 
The RKD cases allow further investigation into the effects of modifying the suprathermal tail through the parameters $\kappa$ and $\alpha$, revealing their interplay in defining plasma stability. 

\subsection{EMEC Instability with RKD Halo}\label{Sec:EMEC_RKD}

\begin{figure*}[t!]
    \centering
    \includegraphics[width=1.0\textwidth]{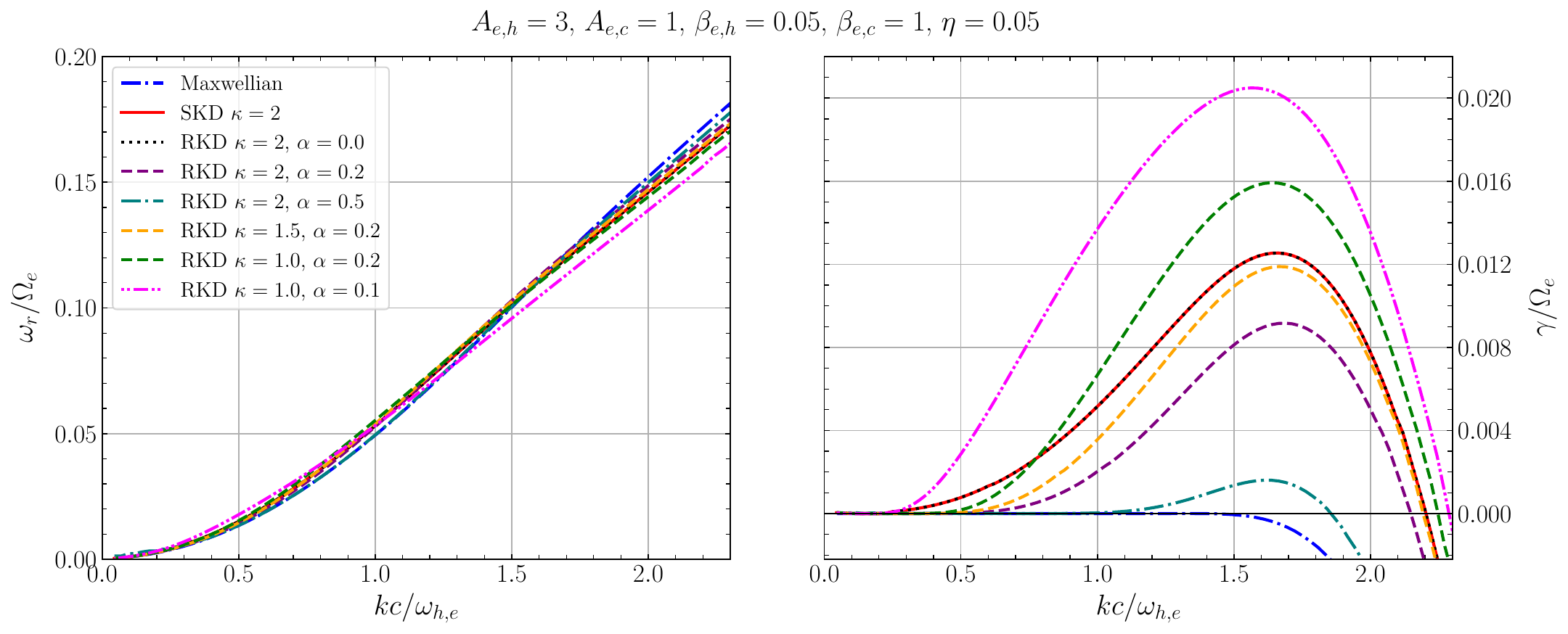}
    \caption{\textit{The EMEC instability with different RKD halos with reference to \cite{Lazar_2017_AA}, case 1. RKDs with the same $\alpha$ are plotted in the same dashed style, while the Maxwellian and Maxwellian-like RKD are dashed-dotted. The RKD with no cut-off (dotted) leads to the same result as the SKD and approaches the Maxwellain results with an increasing cut-off, exhibiting a clear ordering in $\alpha$ regarding the maximum growth rate. }} \label{EMEC_Halo_RKD}
\end{figure*}

The results for the first case ($A_{e,h}=3.0$, $\beta_{e,h}=0.05$) are shown in Fig. \ref{EMEC_Halo_RKD}, with frequency in the panel on the left-hand side and growth rate in the panel on the right-hand side. RKDs with the same $\alpha$ are plotted in the same dashed style, while the Maxwellian and Maxwellian-like curves are dashed-dotted.

First, the RKD with $\kappa=2$ and $\alpha=0$ (dotted black) lead to the same results as the SKD with $\kappa=2$ (solid red), since there is no cut-off of the RKDs suprathermal tails. Both curves, in frequency and growth rate, agree very well, validating the implementing of the RKD in ALPS.
While the frequency is not expected to vary strongly for the different VDFs, ($\omega_r/\Omega_e$ generally increases with the wave number $k d_e$), the choice of the latter should impact the growth rates noticeably.
The RKD distributions exhibit varying growth rates depending on the values of $\kappa$ and $\alpha$. \textcolor{black}{As $\alpha$ increases (and keeping $\kappa$ constant), resulting in a more Maxwellian-like distribution}, the overall growth rates decrease (with a lower maximum, shifted to higher wave numbers), indicating reduced instability due to the less prominent high energy tails.
So as expected, \textcolor{black}{the RKD approaches the Maxwellian results with an increasing cut-off}, indicating a clear ordering in $\alpha$ regarding the maximum growth rate.   
Conversely, lower $\alpha$ values, especially when combined with lower $\kappa$ (e.g., $\kappa = 1.0$), lead to higher growth rates, showing that the distribution becomes more nonthermal and thus more likely to be unstable.

The cut-off parameter can act as a modulation parameter: The RDK with $\kappa = 1.5$ and $\alpha = 0.2$ (yellow dotted) has obviously a larger suprathermal population than the RKDs with $\kappa = 2$ and $\alpha \leq 0.2$, leading to a higher growth rate compared to the latter ones. However, the RDK with $\kappa = 1.5$ and $\alpha = 0.2$ still results in a lower maximum growth rate than the RKD with $\kappa = 2.0$ and no cut-off.
When using an even higher $\kappa = 1.0$ (green dashed), the high energy tails, even with a cut-off $\alpha = 0.2$, are dominant enough to exhibit a much higher growth rate, and even more so for $\kappa = 1.0$  with $\alpha = 0.1$ (double-dotted dashed, magenta), when the growth rate evolves into a much wider peak, illustrating enhanced instability over an extended range of wave numbers reflecting the combined impact of a minimal cut-off and strong suprathermal presence. 

This pattern signifies, as expected, that reducing $\kappa$ (increasing the suprathermal component) without increasing $\alpha$ (keeping the cut-off relatively weak) results in a more destabilizing effect. Both of the results with high growth rates would not be achievable when using a SKD.
The mentioned plots of the VDFs, Fig. \ref{plot_Halo_VDF_ALPSlike} and Fig. \ref{contourplot_Halo_VDF_normalized}, are in plausible agreement with the obtained results, i.e.: The density of the halo component for $\kappa = 2.0$ and $\alpha = 0.0$ is higher at high velocities than for $\kappa = 1.5$ and $\alpha = 0.2$, resulting in a higher growth rate for the former. 

\begin{figure*}[t!]
    \centering
    \includegraphics[width=1.0\textwidth]{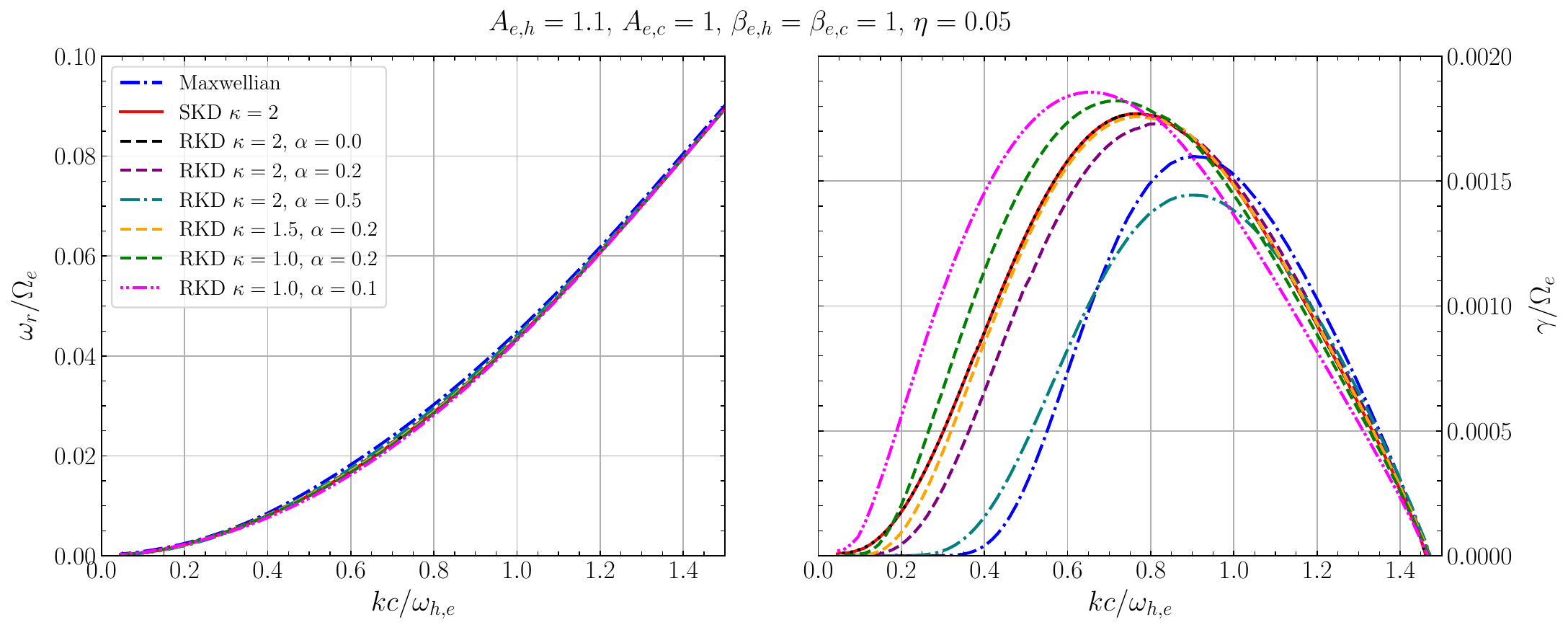}
    \caption{\textit{The EMEC instability with different RKD halos with reference to \cite{Lazar_2017_AA}, case 2. RKDs with the same $\alpha$ are plotted in the same dashed style, while the Maxwellian and Maxwellian-like RKD are dashed-dotted. The RKD with no cut-off (dotted) leads to the same result as the SKD and approaches the Maxwellain results with an increasing cut-off, exhibiting a clear ordering in $\alpha$ regarding the maximum growth rate. }} \label{Fig:EMEC_Halo_RKD_lowA}
\end{figure*}

The results for the second case ($A_{e,h}=1.1$, $\beta_h=1$) are shown in Fig. \ref{Fig:EMEC_Halo_RKD_lowA}. As expected, the frequency hardly varies between the different VDFs. Again, the results for the RKD with $\kappa = 2$ and no cut-off $\alpha = 0$ (dotted black) are in very good agreement with the SKDs results.
When $\alpha$ is increased to $0.2$ (dashed purple) the peak growth rate shifts toward larger wave vectors, with the maximum value slightly decreasing. This reduction in growth rate and shift in peak location signals a dampening effect due to the exponential cut-off, which begins to moderate the influence of the suprathermal tail. For $\alpha = 0.5$  (dash-dotted teal), this effect is even more pronounced, as the maximum growth rate peak shifts to higher wavenumbers, comparable to the Maxwellian case, and the maximum growth rate diminishes further, even below the the Maxwellian case. 
Similar to the first case, when $\kappa$ is decreased to $1.5$, but with a cut-off $\alpha = 0.2$ (dashed orange), the growth rate is very similar to the SKD/RKD without cut-off, with differences mostly at low wavenumbers.  
For the RKD cases with $\kappa = 1.0$ (dashed green and double-dotted dashed magenta), which represent distributions with the strongest suprathermal effects, the growth rate for both $\alpha$ cases exhibits a higher maximum growth rate at low values of $k$. Notably, the $\alpha = 0.1$ case achieves the highest peak growth rate.

Following \cite{Husidic-etal-2022}, we introduce the ratios $R_\gamma = \gamma_{\text{max}, i}/\gamma_{\text{max}, \text{SKD}}$ and $R_k = k_{\text{max}, i}/k_{\text{max}, \text{SKD}}$ to compare the maximum growth rate and corresponding $k$-value of the SKD with those of the other VDFs $i$. The results are presented in table \ref{tab:EMEC_case1} for the first case and in Table \ref{tab:EMEC_case2} for the second case. The results for $R_\gamma$ show the described and expected behavior in both cases, although more prominent for the case with a higher anisotropy and $R_k$ has also a clear ordering.

\subsection{EFHI with RDK Halo}
\begin{figure*}[t!]
    \centering
    \includegraphics[width=1.0\textwidth]{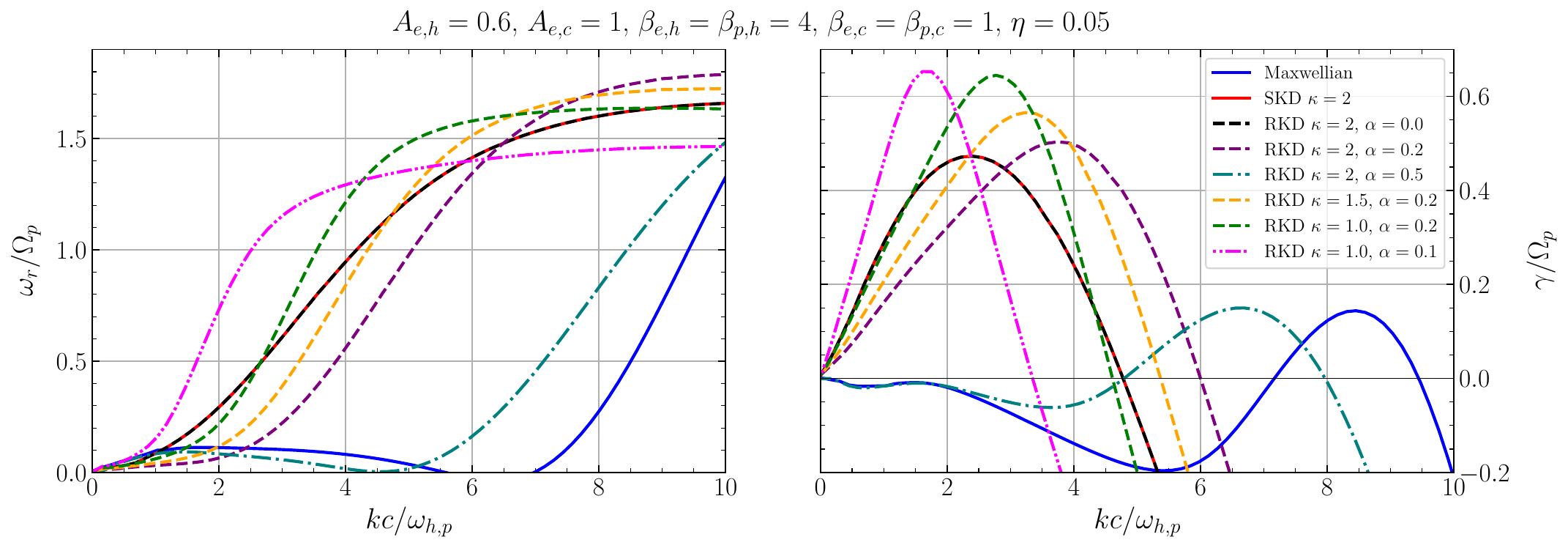}
    \caption{\textit{The EFHI with different RKD Halos with reference to Lazar 2017, case 1. RKDs with the same $\alpha$ are plotted in the same dashed style, while the Maxwellian and Maxwellian-like RKD are dashed-dotted. The RKD with no cut-off (dotted) leads to the same result as the SKD and approaches the Maxwellain results with an increasing cut-off, exhibiting a clear ordering in $\alpha$ regrading the maximum growth rate. }} \label{Fig:EFHI_Halo_RKD}
\end{figure*}

\begin{figure*}[t!]
    \centering
    \includegraphics[width=0.45\textwidth]{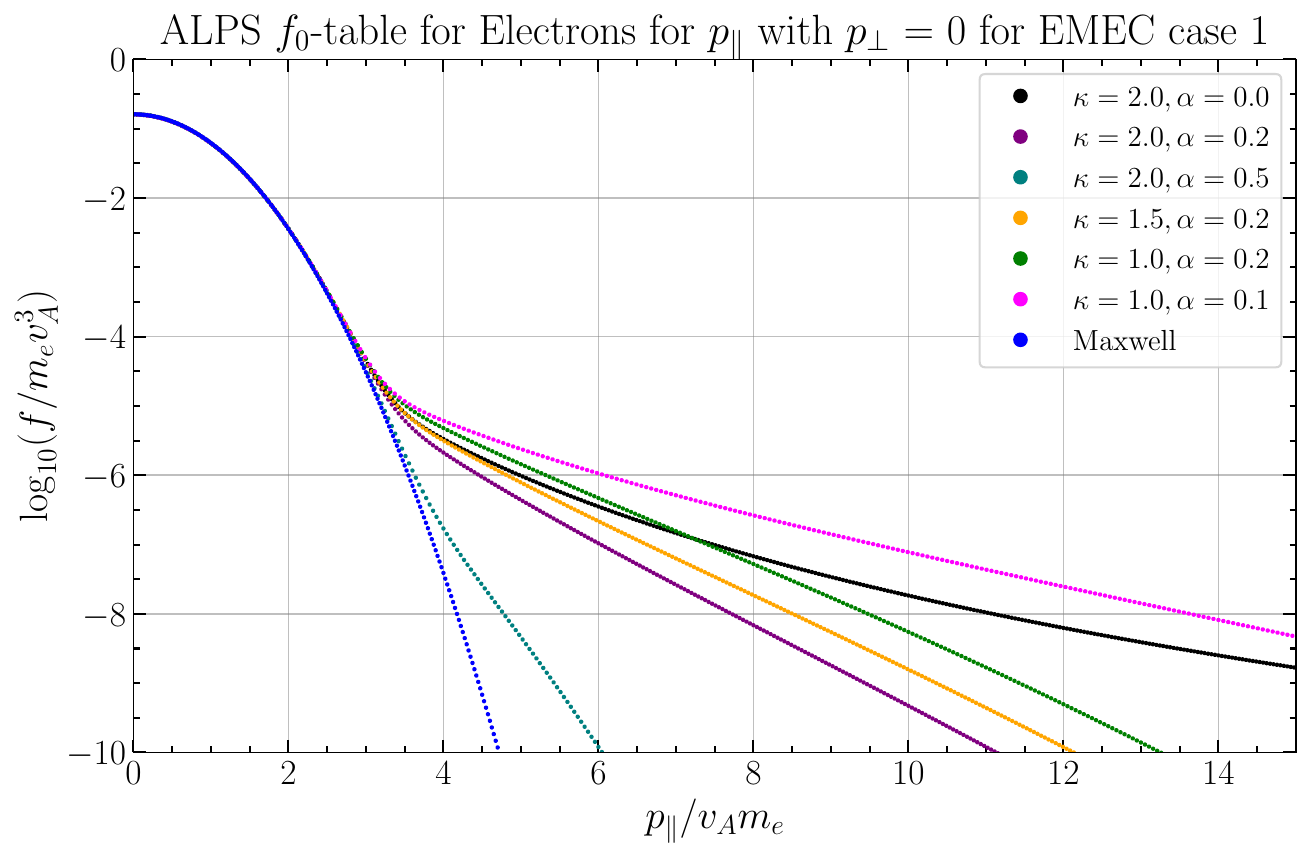}
    \caption{\textit{A comparison between the different anisotopic Core+Halo VDFs for EMEC case 1. This represents the $f_0$ tables that are used in ALPS, the dots correspond to actual data points.  }} \label{plot_Halo_VDF_ALPSlike}
\end{figure*}

The EFHI is also extended to cases with different RKD halos, using the same combination of $\kappa$ and $\alpha$ as for the EMEC cases. The results are shown in Fig \ref{Fig:EFHI_Halo_RKD}, with the frequency in the panel on the left-hand side and the growth rate in the panel on the right-hand side. The structure of the chosen plot styles is the same as for the EMEC, the results for the calculated values of $R_{\gamma}$ and $R_{k}$ can be obtained form Table \ref{tab:EFHI_case1}. The VDF with $\kappa = 2$ and no cut-off $\alpha = 0$ leads to the same results as the SKD case, as expected.
Increasing $\alpha$ to $0.2$ introduces a subtle flattening of the increase of the frequency for low wave numbers and shifts the maximum growth rate to higher values of $k$. While the shape of the peak broadens, the maximum growth rate increases.  With a further increase to $\alpha = 0.5$ (dash-dotted teal), the frequency profile is closely approaching the behavior of the Maxwellian case. The growth rate in this case exhibits a low maximum value, comparable to the Maxwellian case, just shifted to a lower value of $k$. The overall shape is also similar to the Maxwellian case, with near identity for $kc/\omega_{h,p}<2$, suggesting significant stabilization in the presence of a strong exponential cut-off.
When $\kappa$ is decreased to $1.5$, while keeping the cut-off at $\alpha = 0.2$, the maximum growth rate increases and is shifted to a higher wave number, compared to the SKD/RKD no cut-off case. Decreasing $\kappa$ even further to $1.0$, with the same $\alpha = 0.2 $, leads to a noticeable higher maximum growth rate, with a more prominent, sharper peak and with only a slight shift to higher $k$ values. For a smaller $\alpha = 0.1$, the frequency shows sharp initial rises before quickly plateauing, compared to the SKD/RKD no cut-off results. The maximum growth rate is, compared to the previously case with higher $\alpha = 0.2$, only slightly increased. The peak however, is much narrower at lower wave numbers (in comparison to the SKD/RKD no cut-off case), covering a much smaller range of $k$ values.  
Overall the RKD halo model (e.g., Fig. \ref{plot_Halo_VDF_ALPSlike}) effectively illustrates the substantial impact that varying suprathermal components have on the EFHI, underline the flexibility of RKD models in capturing diverse plasma environments, with even highly nonthermal components, and provides a useful framework for exploring these instabilities.  

\begin{table*}[t]
    \centering
    \begin{tabular}{@{\hspace{10pt}} c @{\hspace{10pt}} c @{\hspace{10pt}} c @{\hspace{10pt}} c @{\hspace{10pt}}}
        \hline
        $\kappa$ & $\alpha$ & $R_\gamma$ & $R_k$ \\
        \hline
        $\infty$ & - & 0.0 & 0.0 \\
        2.0 & 0.5 & 0.128 & 0.970 \\
        2.0 & 0.2 & 0.731 & 1.016 \\
        2.0 & 0.0 & 1.0 & 1.0 \\
        1.5 & 0.2 & 0.947 & 1.0 \\
        1.0 & 0.2 & 1.269 & 0.985 \\
        1.0 & 0.1 & 1.633 & 0.939\\
        \hline
    \end{tabular}
    \caption{\textit{Comparison of maximum growth rate and corresponding wave numbers (both normalized to the values of the SKD) for the EMEC case 1.  }}
    \label{tab:EMEC_case1}
\end{table*}

\begin{table*}[t]
    \centering
    \begin{tabular}{@{\hspace{10pt}} c @{\hspace{10pt}} c @{\hspace{10pt}} c @{\hspace{10pt}} c @{\hspace{10pt}}}
        \hline
        $\kappa$ & $\alpha$ & $R_\gamma$ & $R_k$ \\
        \hline
         $\infty$ & - & 0.899 & 1.185 \\
         2.0 & 0.5 & 0.816 & 1.173 \\
         2.0 & 0.2 & 0.977 & 1.108 \\
         2.0 & 0.0 & 1.0 & 1.0 \\
         1.5 & 0.2 & 0.993 & 1.021 \\
         1.0 & 0.2 & 1.029 & 0.933 \\
         1.0 & 0.1 & 1.049 & 0.868 \\
        \hline
    \end{tabular}
    \caption{\textit{Comparison of maximum growth rate and corresponding wave numbers (both normalized to the values of the SKD) for the EMEC case 2.}}
    \label{tab:EMEC_case2}
\end{table*}

\begin{table*}[t]
    \centering
    \begin{tabular}{@{\hspace{10pt}} c @{\hspace{10pt}} c @{\hspace{10pt}} c @{\hspace{10pt}} c @{\hspace{10pt}}}
        \hline
        $\kappa$ & $\alpha$ & $R_\gamma$ & $R_k$ \\
        \hline
         $\infty$ & - & 0.305 & 3.545 \\
         2.0 & 0.5 & 0.317 & 2.807 \\
         2.0 & 0.2 & 1.064 & 1.569 \\
         2.0 & 0.0 & 1.0 & 1.0 \\
         1.5 & 0.2 & 1.196 & 1.355 \\
         1.0 & 0.2 & 1.363 & 1.163 \\
         1.0 & 0.1 & 1.380 & 0.675 \\
        \hline
    \end{tabular}
    \caption{\textit{Comparison of maximum growth rate and corresponding wave numbers (both normalized to the values of the SKD) for the EFHI case 1.}}
    \label{tab:EFHI_case1}
\end{table*}




\section{Summary}\label{sec:conclusion}
We study the dispersion relation and linear instability of plasma systems with a background distribution function that follows a regularized
bi-$\kappa$-halo model.
 Low $\kappa$ values, which enhance the suprathermal population, significantly increase growth rates and hence amplify instability. Conversely, higher $\alpha$ values have a stabilizing influence by lowering the suprathermal tail, shifting the behavior toward that of the Maxwellian or SKD and reducing the maximum growth rates and instability. These findings underline the importance of core-halo characteristics in determining plasma wave stability.

 A $\kappa < 3/2$ and moderate $\alpha$ significantly enhance wave instability. This parameter combination would not be accessible for a plasma representation with an SKD. These findings are significant for understanding the conditions under which EMEC waves and EFHI waves become unstable in space and astrophysical plasmas. The enhanced instability for RKD distributions with  $\kappa  < 3/2$ and lower $\alpha$ values are particularly relevant for environments with prevalent nonthermal populations, such as the solar wind. Furthermore, the cut-off parameter $\alpha$ can be used as a modulation parameter, which is not present in the SKD and suggests that RKDs are a better model for capturing the effects of nonthermal populations in certain plasma environments. The study also demonstrated that ALPS is a powerful and versatile numerical tool to investigate instabilities of arbitrary VDFs, justifying its use for further studies.



\begin{acknowledgments}
The authors acknowledge support from the Ruhr-University Bochum, the Katholieke Universiteit Leuven, and the use of the ALPS code. This project was funded by the Deutsche Forschungsgesellschaft (DFG), project FI $706/31$-$1$ and the Belgian FWO-Vlaanderen G002523N, and SIDC Data Exploitation (ESA Prodex), No. 4000145223. The ALPS project received support from UCL’s Advanced Research Computing Centre through the Open Source Software Sustainability Funding scheme. D.Verscharen is supported by STFC Consolidated Grant ST/W001004/1. K.G. Klein was supported by NASA ECIP Grant 80NSSC19K0912.

\end{acknowledgments}
\section*{Data Availability Statement}
 The ALPS code is available via an open source BSD 2-Clause License at \url{https://github.com/danielver02/ALPS} with a full tutorial on its use at \url{https://danielver02.github.io/ALPS/}. The data that support the findings of this study are available from
the corresponding author upon reasonable request.
\newpage
\appendix

\
\
\
\section{Dispersion relation in ALPS} \label{app_disper}
The susceptibilities can be expressed via
\begin{widetext}
\begin{equation}
\boldsymbol{\chi}_j=\frac{\omega_{\mathrm{p} j}^2}{\omega \Omega_{j}} \int_0^{\infty} 2 \pi p_{\perp} \mathrm{d} p_{\perp} \int_{-\infty}^{+\infty} \mathrm{d} p_{\|}\left[\hat{\boldsymbol{e}}_{\|} \hat{\boldsymbol{e}}_{\|} \frac{\Omega_{j}}{\omega}\left(\frac{1}{p_{\|}} \frac{\partial f_{0 j}}{\partial p_{\|}}-\frac{1}{p_{\perp}} \frac{\partial f_{0 j}}{\partial p_{\perp}}\right) p_{\|}^2\right. \\
\left.+\sum_{n=-\infty}^{+\infty} \frac{\Omega_{j} p_{\perp} U}{\omega-k_{\|} v_{\|}-n \Omega_{j}} \boldsymbol{T}_n\right]
\end{equation}
\end{widetext}
Here, $\omega_{\mathrm{p} j} \equiv \sqrt{4 \pi n_j q_j^2 / m_j}$ is the plasma frequency of the species. The tensor $\boldsymbol{T}_n$ is defined as
$$
\boldsymbol{T}_n \equiv\left(\begin{array}{ccc}
\frac{n^2 J_n^2}{z^2} & \frac{i n J_n J_n^{\prime}}{z} & \frac{n J_n^2 p_{\|}}{z p_{\perp}} \\
-\frac{i n J_n J_n^{\prime}}{z} & \left(J_n^{\prime}\right)^2 & -\frac{i J_n J_n^{\prime} p_{\|}}{p_{\perp}} \\
\frac{n J_n^2 p_{\|}}{z p_{\perp}} & \frac{i J_n J_n^{\prime} p_{\|}}{p_{\perp}} & \frac{J_n^2 p_{\|}^2}{p_{\perp}^2}
\end{array}\right),$$
with $z \equiv k_{\perp} v_{\perp} / \Omega_{j}$, and $J_n \equiv J_n(z)$ as the $n$ th-order Bessel function. And 
\begin{equation}
U \equiv \frac{\partial f_{0 j}}{\partial p_{\perp}}+\frac{k_{\|}}{\omega}\left(v_{\perp} \frac{\partial f_{0 j}}{\partial p_{\|}}-v_{\|} \frac{\partial f_{0 j}}{\partial p_{\perp}}\right)
\end{equation}
\
\

\section{Dispersion relation for dual Maxwellian-$\kappa$-model}
The dispersion relations for the dual Maxwellian-$\kappa$-model were derived in \cite{Lazar_2017_AA} and are solved with Mathematica for this paper.
\subsection{EMEC case} 
For the EMEC instability, the dispersion relation reads 
\begin{equation}\label{eq:disp_emec}
\begin{aligned}
(kc/\omega_{h,e})^2=A_{e, h}-1+ & \frac{A_{e, h}(\omega/|\Omega_e|-1)+1}{kc/\omega_{h,e} \sqrt{a^2 \beta_{e, h}}} Z_{\kappa}\left(\frac{\omega/|\Omega_e|-1}{kc/\omega_{h,e} \sqrt{a^2 \beta_{e, h}}}\right) \\
& +\frac{1}{\eta}\left[\frac{\omega/|\Omega_e|}{kc/\omega_{h,e} \sqrt{\eta \beta_{e, c}}} Z_M\left(\frac{\omega/|\Omega_e|-1}{kc/\omega_{h,e} \sqrt{\eta \beta_{e, c}}}\right)\right]
\end{aligned}
\end{equation}
with $a=(1-1.5/\kappa)^{0.5}$ and the plasma dispersion function $Z_M$ for the Maxwellian case 
\begin{equation}
\begin{aligned}
Z_{j, M}\left(\xi_{j, M}^{ \pm}\right)=\frac{1}{\pi^{1 / 2}} \int_{-\infty}^{\infty} \frac{\exp \left(-x^2\right)}{x-\xi_{j, M}^{ \pm}} d t, \quad \mathfrak{J}\left(\xi_{j, M}^{ \pm}\right)>0
\end{aligned}
\end{equation}
with the argument $\xi_{j, M}^{ \pm}=(\omega\pm\Omega _j)/(k\theta_{j,\|})$, where $\pm$ denotes the circular polarization and the plasma dispersion function  $Z_{\kappa}$ for the SKD
\begin{equation}
\begin{aligned}
Z_{j, \kappa}\left(\xi_{j, k}^{ \pm}\right) & =\frac{1}{\pi^{1 / 2} \kappa^{1 / 2}} \frac{\Gamma(\kappa)}{\Gamma(\kappa-1 / 2)} \\
& \times \int_{-\infty}^{\infty} \frac{\left(1+x^2 / \kappa\right)^{-\kappa}}{x-\xi_{j, \kappa}^{ \pm}} d x, \mathfrak{J}\left(\xi_{j, k}^{ \pm}\right)>0,
\end{aligned}
\end{equation}
with the argument $\xi_{j, \kappa}^{ \pm}=(\omega\pm\Omega _j)/(k\theta_{j,\|})$.
\subsection{EFHI case}
For the EFHI, the dispersion relation reads
\begin{equation}\label{eq:disp_efhi}
\begin{array}{r}
(kc/\omega_{h,p})^2=\mu\left[A_{e, h}-1+\frac{A_{e, h}(\omega/\Omega_p+\mu)-\mu}{kc/\omega_{h,p}\sqrt{\alpha^2 \mu \beta_{e, h}}} Z_\kappa\left(\frac{\omega/\Omega_p+\mu}{kc/\omega_{h,p} \sqrt{\alpha^2 \mu \beta_{e, h}}}\right)\right] \\
+\frac{1}{\eta}\left[\frac{\omega/\Omega_p}{kc/\omega_{h,p}\sqrt{\eta \beta_{p, c}}} Z_M\left(\frac{\omega/\Omega_p-1}{kc/\omega_{h,p}\sqrt{\eta \beta_{p, c}}}\right)\right]\\+ \frac{\omega/\Omega_p}{kc/\omega_{h,p}\sqrt{\alpha^2 \beta_{p, h}}} Z_\kappa\left(\frac{\omega/\Omega_p-1}{kc/\omega_{h,p}\sqrt{\alpha^2 \beta_{p, h}}}\right) \\
+\frac{\mu}{\eta}\left[\frac{\omega/\Omega_p}{kc/\omega_{h,p} \sqrt{\mu \eta \beta_{e, c}}} Z_M\left(\frac{\omega/\Omega_p+\mu}{kc/\omega_{h,p} \sqrt{\mu \eta \beta_{e, c}}}\right)\right]
\end{array}
\end{equation}
with $\mu=m_p/m_e=1836$ and the other parameters as above.

\nocite{*}
\bibliography{aipsamp}

\end{document}